\documentclass[prl,aps,twocolumn,showpacs,preprintnumbers,amsmath,amssymb]{revtex4}
\usepackage{bm}
\usepackage{epsfig}

\begin{document}

\preprint{}

\title{3:1 magnetization plateau and suppression of ferroelectric polarization in an Ising chain multiferroic}

\author{Y. J. Jo,$^1$  Seongsu Lee,$^2$ E. S. Choi,$^1$ H. T. Yi,$^2$, W.
Ratcliff II,$^3$, Y. J. Choi,$^2$   V. Kiryukhin,$^2$ S. W.
Cheong,$^2$ and L. Balicas,$^1$}

\affiliation{$^1$National High Magnetic Field Laboratory, Florida
State University, Tallahassee-FL 32310, USA} \affiliation{$^2$
Rutgers Center of Emergent Materials and Department of Physics \&
Astronomy, Rutgers University, Piscataway, New Jersey, 08854, USA}
\affiliation{$^3$ NIST Center for Neutron Research, National
Institute of Standards and Technology, Gaithersburg, Maryland
20899, USA}

\date{\today}%

\begin{abstract}

Ferroelectric Ising chain magnet Ca$_3$Co$_{2-x}$Mn$_x$O$_6$
($x\simeq$0.96) was studied in magnetic fields up to 33 T.
Magnetization and neutron scattering measurements reveal
successive metamagnetic transitions from the zero-field $\uparrow
\uparrow \downarrow \downarrow$ spin configuration to the
$\uparrow \uparrow \uparrow \downarrow$ state with a broad
magnetization plateau, and then to the $\uparrow \uparrow \uparrow
\uparrow$ state. The absence of hysteresis in these plateaus
reveals an intriguing coupling between the intra-chain state and
the three-dimensional geometrically frustrated magnetic system.
Inversion symmetry, broken in the $\uparrow \uparrow \downarrow
\downarrow$ state, is restored in the $\uparrow \uparrow \uparrow
\downarrow$ state, leading to the complete suppression of the
electric polarization driven by symmetric superexchange.

\end{abstract}

 \maketitle

Recent discoveries of new multiferroics (compounds exhibiting both
magnetism and ferroelectricity) that display a coupling between
the corresponding order parameters has triggered a resurgence in
the field of the magnetoelectric effect
\cite{cheong,khomski,fiebig,mathur}. In many of these materials,
magnetic order breaks inversion symmetry, and electric
polarization is subsequently induced via magnetoelastic coupling.
Typically, such magnetism-driven ferroelectricity is found in
spiral magnets, in which the spin-lattice coupling arises due to
the Dzyaloshinskii-Moriya interaction, that is associated with the
antisymmetric part of the exchange coupling \cite{cheong}. This
mechanism is realized in TbMnO$_3$, Ni$_3$V$_2$O$_8$, CuFeO$_2$,
and in many other systems \cite{cheong,kimura,lawes,kimura2}.
However, because of the weakness of the antisymmetric exchange,
the induced electric polarization in these systems is rather
small. Exchange striction associated with symmetric superexchange
coupling is, in general, a considerably larger effect. It may give
rise to giant atomic displacements, such as those recently
observed in hexagonal (YLu)MnO$_3$ \cite{lee}, and thereby affect
dielectric properties. Multiferroics with symmetric exchange
striction therefore hold significant promise as candidate
materials for giant magnetoelectric effects.

A remarkable example of the latter system is a recently discovered
multiferroic Ising chain magnet Ca$_3$Co$_{2-x}$Mn$_x$O$_6$ (for
$x$ $\approx$0.96) \cite{choi}. It is composed of $c$-axis spin
chains consisting of magnetic ions within alternating oxygen cages
of face-shared trigonal prisms and octahedra. The spin chains are
separated by the Ca ions and form a triangular lattice within the
$ab$-plane. For $x=1$, all Co ions are located in the prismatic
sites, while all the Mn ions occupy the octahedral sites
\cite{subkov}. Because of the frustrated ferromagnetic nearest
neighbor (nn) and antiferromagnetic next nearest neighbor (nnn)
interactions within the chains, an $\uparrow \uparrow \downarrow
\downarrow$ magnetic order forms for $T<T_c\approx$16 K
\cite{choi}. Combined with the alternating Co$^{2+}$ and Mn$^{4+}$
ionic order, this breaks the inversion symmetry and induces an
electric polarization along the chain via symmetric
magnetostriction \cite{cheong,choi}. Thus, this system combines
physics of frustrated Ising chain with physics of
magnetically-driven ferroelectricity. Frustrated Ising chains were
extensively studied theoretically \cite{selke,jjkim}, but good
experimental realizations were proven hard to find. In addition,
because of the triangular coordination of the chains within the
$ab$ plane, the interchain interaction is geometrically
frustrated. Effects of frustration are revealed in this material
by the observation of a magnetic freezing transition at
$T_F\approx$3 K \cite{choi}. Here we explore the effects of the
application of an external magnetic field on the magnetic and
ferroelectric response of Ca$_3$Co$_{2-x}$Mn$_x$O$_6$ (for $x$
$\approx$0.96). This is of great interest given the complex
interplay between the physics of frustrated Ising spin chains, the
magnetoelectric effect, and the geometrically frustrated
magnetism. Below, we show that Ca$_3$Co$_{2-x}$Mn$_x$O$_6$
provides a remarkable model to explore the interrelation among
these phenomena.

\begin{figure}[htb]
\begin{center}
\epsfig{file= 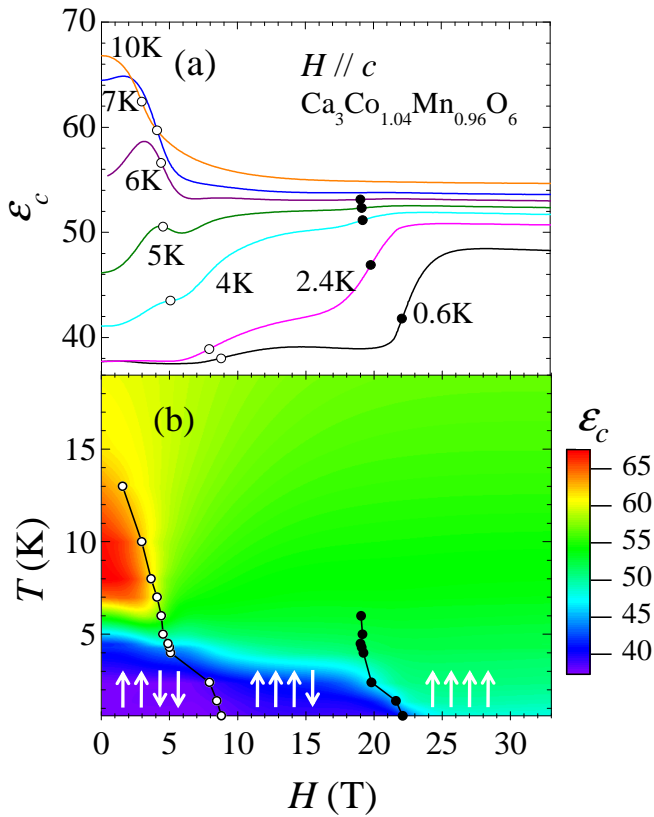, width = 8.6 cm}
\caption{(color online)(a) Dielectric constant $\varepsilon_c$ for
a Ca$_3$Co$_{1.04}$Mn$_{0.96}$O$_6$ single crystal for an electric
field applied along the $c$ axis and as a function of the magnetic
field $H$ applied in the same direction. (b) Contour plot
displaying the behavior of $\varepsilon_{c}$ in the $T$-$H$ plane.
Circles indicate magnetic transitions. Spin configurations are
shown with arrows (see text). Open and closed circles correspond in both figures to the maximums in
the derivative of $\varepsilon_c$ respect to the field which are
associated with the field-induced phase transitions.}
\end{center}
\end{figure}

Single crystals of Ca$_{3}$Co$_{1.04}$Mn$_{0.96}$O$_{6}$ were
grown by the flux method described in Ref. \cite{choi} producing
needle-like single crystals. These were cut in a convenient
geometry for dielectric constant $\varepsilon$ and electric
polarization $P$ measurements, with a cross-sectional area of
0.45-0.55 mm$^2$ and thickness of 0.32-0.35 mm. To remove any
strain left by the cutting/polishing procedure, the crystals were
annealed at 650$^{\circ}$C for 5 hours. Two electrodes were
painted onto the largest surfaces of the crystals. $\varepsilon$
was obtained by measuring the capacitance as a function of
temperature or field with a manual capacitance bridge, using an
excitation signal of 30 V at 5 kHz. To obtain the temperature and
the magnetic field dependence of the electric polarization $P$ the
pyroelectric and magnetoelectric currents were measured with an
electrometer at a rate of respectively, 3 K/min and 5T/min, after
cooling the specimens from 40 to 1.5 K in a static poling electric
field $E_{pole}$=5.63 kV/m. $P(T, H)$ was measured after removing
the poling field. Magnetization measurements were performed by
using a Vibrating Sample Magnetometer in magnetic fields up to 33
T coupled to a $^3$He refrigerator provided by the NHMFL. Neutron
powder diffraction measurements were performed in polycrystalline
samples of Ca$_3$Co$_{1.05}$Mn$_{0.95}$O$_6$ at the BT-7 beam line
at NIST Center for Neutron Research. The specimen was loaded in a
11 tesla vertical field magnet. The data was collected at $T =
1.6$ K in the horizontal scattering plane by using monochromatic
neutrons of wavelength 2.3592 \AA. The diffraction data was
refined using the FULLPROF program package \cite{fullprof}. Due to
strong magneto-crystalline anisotropy, the crystallites tend to
orient along the $c$-axis in applied magnetic fields. 
\begin{figure}[htb]
\begin{center} \epsfig{file=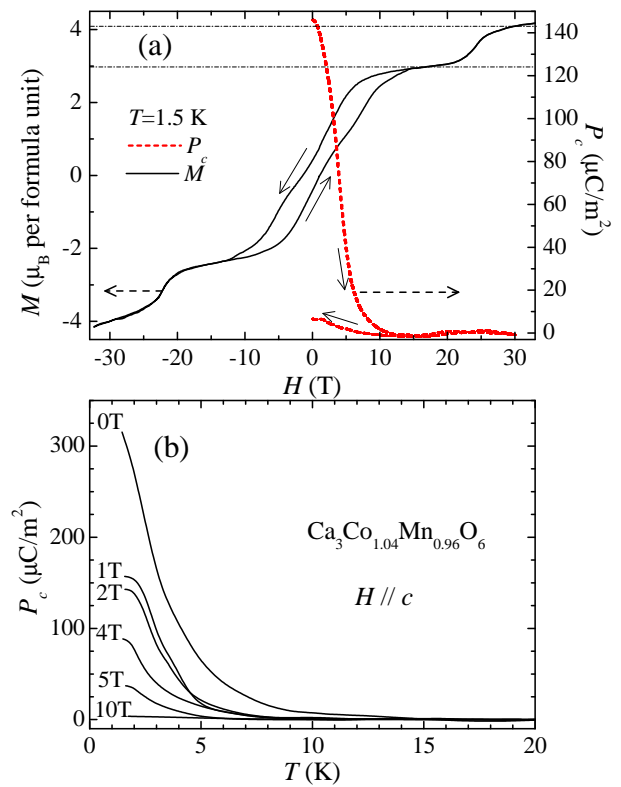,width=8 cm}
\caption {(color online) (a) Magnetization $M$, and $c$-axis electric
polarization $P_c$ as functions of the field $H$ applied along the
$c$ axis at $T$=1.5 K and 1.4 K, respectively. Notice the marked hysteresis 
on both $M$ and $P_c$ among field up and down sweeps. (b) Electric
polarization $P_c$ as a function of temperature for several values
of $c$-axis magnetic field.}
\end{center}
\end{figure}
A correction
for this effect (preferred orientation), obtained from refinement
of the nuclear structure only, was applied in the final refinement
of the magnetic and nuclear structures.

Figure 1(a) shows the magnetic field dependence of the dielectric
constant $\varepsilon_c$ for an electric field along the chain
direction, i.e. $E
\parallel c$, and for several temperatures. The magnetic field was also
applied along the chain direction. As seen, $\varepsilon_c$
displays broad temperature dependent features: a broad peak below
5 T that is displaced to higher fields as $T$ is decreased, and a
step for $20 \leq H \leq 25 $ T. As is shown below, both features
coincide with field-induced magnetic and ferroelectric
transitions. The overall dependence of $\varepsilon_c$ on $H$ and
$T$ is displayed in Fig. 1 (b), which shows a contour plot of
$\varepsilon_c$ in the $T-H$ plane. $\varepsilon_c$ shows a broad
maximum around $\sim 10$ K, which is several degrees below the
magnetic phase transition to the the $\uparrow \uparrow \downarrow
\downarrow$ spin state, with all the spins pointing along the
c-axis \cite{choi}. The discrepancy between the temperatures of
the magnetic transition and the maximum in $\varepsilon_c$ results
from the freezing effects and corresponding quasi-long-range
magnetic order \cite{choi}. As shown in Ref. \cite{choi} these
effects also cause the magnetoelectric freezing transition for
$T<$5 K, which is revealed in the behavior of the magnetic
susceptibility, as well as in the marked suppression of
$\varepsilon_c$ clearly observed in the data of Fig. 1(b). 

\begin{figure}[htb]
\begin{center}
\epsfig{file=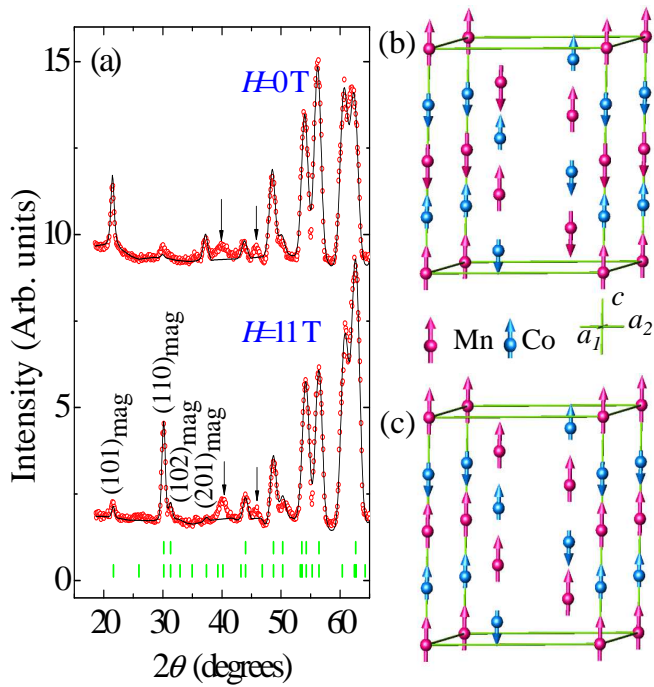, width=8.6 cm} \caption{(color
online) (a) Observed (symbols) and calculated (line) powder
neutron diffraction patterns for $H$=0 and 11 T, both at $T$=1.6
K. The bars below the patterns indicate positions for nuclear
(first row) and magnetic (second row) Bragg peaks. Arrows show the
impurity phase. (b) Zero-field magnetic $\uparrow \uparrow
\downarrow \downarrow$ structure \cite{choi}. (c) The $\uparrow
\uparrow \uparrow \downarrow$ magnetic structure obtained in the
refinement of the 11 T data.}
\end{center}
\end{figure}

To elucidate the origin of the features observed in
$\varepsilon_c$, we performed magnetization ($M$) measurements as
a function of magnetic field applied along the chain direction. A
representative curve is shown in Fig. 2 (a) for $T$=1.5 K. Two
major features are observed in $M(H)$, largely coinciding with the
structures seen in $\varepsilon_c$. The first feature at
approximately 10 T leads to a very broad plateau at a value of
$\sim 3$ $\mu_B$ corresponding to the full saturation moment of
Mn$^{4+}$, thus indicating that the Mn sublattice becomes
ferromagnetic above this field. A second quasi-plateau is observed
above 25 T with a saturation moment of $\sim 4$ $\mu_B$,
corresponding to a fully polarized spin state of high-spin
Mn$^{4+}$ and low-spin Co$^{2+}$ moments with no apparent large
orbital contribution found in neutron diffraction experiments
\cite{choi}. Thus $M$ indicates that the $\uparrow \uparrow
\uparrow \downarrow$ spin state with \emph{all} Mn$^{4+}$ moments
aligned along the field is stabilized for $10 \leq H \leq 25$ T,
and that the system becomes ferromagnetic at higher fields.
Importantly, this sequence of transitions is predicted by the
alternating-spin Ising ferrimagnet model with competing nn and nnn
interactions in a large parameter range \cite{jjkim}. It is
therefore probable that such a simple model is physically realized
in our system. Finally, in our single crystals, geometrical
frustration leads to magnetic domain formation at $H$=0
\cite{choi}. This would explain the hysteretic behavior seen in
$M$ for $0 \leq H \leq 15$ T. Intriguingly, it disappears once the
$\uparrow \uparrow \uparrow \downarrow$ state is stabilized. The
absence of hysteresis in the plateau region suggests that the
inter-chain geometric frustration has somehow been lifted.
Therefore, a transition in the 1D magnetic subsystem seems to
relieve the geometrical frustration in the 3D lattice, an aspect
that clearly deserves theoretical attention.

Measurements of the electrical polarization $P_c$ as a function of
both temperature and magnetic field applied along the chain
direction reveal a giant magnetoelectric effect. In Fig. 2(a) we
show the electrical polarization $P_c$ as a function of $H$ at a
temperature $T = 1.4$ K. Application of an external field quickly
suppresses $P_c(T)$ which disappears completely for fields
exceeding $\sim 10$ T, i.e, as the $\uparrow \uparrow \uparrow
\downarrow$ plateau is stabilized in $M$. Figure 2 (b) shows the
temperature dependence of $P_c$ for several values of the external
field indicated in the figure. In agreement with the previous
report \cite{choi}, at zero field and below ~15 K, $P_c$ gradually
increases reaching a value of 310 $\mu C/m^2$ at 2 K. The
polarization is completely suppressed above the magnetic
transition fields determined from the $M$ and $\varepsilon_c$
curves shown in Figs. 1 and 2. Thus, our data show that the
field-induced transition from the $\uparrow \uparrow \downarrow
\downarrow$ to the $\uparrow \uparrow \uparrow \downarrow$
structure leads to the total suppression of the electric
polarization. We note that the observed strong magnetoelastic
coupling may play a role in the stabilization of the broad
$\uparrow \uparrow \uparrow \downarrow$ plateau, as it does in the
case of a similar 3:1 plateau observed in $M$Cr$_2$O$_4$ spinels
\cite{ueda}.

To confirm the spin structure suggested by the magnetization
measurements, we performed neutron powder diffraction
measurements. Fig. 3 shows the obtained data at $T$=1.6 K for
fields of zero and 11 T, respectively. The zero-field data
confirms the $\uparrow \uparrow \downarrow \downarrow$ structure
reported in \cite{choi}. In a polycrystalline sample, the
crystallites exhibit all possible orientations with respect to the
external field, and therefore for some of them the $c$-axis
projection of the field is always smaller than the critical field
needed to stabilize the $\uparrow \uparrow \uparrow \downarrow$
state. The critical field may also depend on location in the
multidomain magnetic state. Thus, in an applied magnetic field of
11 T, the sample is expected to contain a mixture of both the
$\uparrow \uparrow \downarrow \downarrow$ and $\uparrow \uparrow
\uparrow \downarrow$ states, with each domain exhibiting its own
spin canting angle with respect to the $c$ axis. The canting angle
averages to zero over the entire sample. The 11 T magnetic
structure was refined assuming the presence of both of structures
with no spin canting. The resulting fit ($R_B$=5.7 \%, $R_M$=12.8
\%) is shown in Fig. 3. The Mn$^{4+}$ moments are 2.20(7) $\mu_B$
and are ordered ferromagnetically, while the Co$^{2+}$ are ordered
antiferromagnetically with a moment of 0.75(8) $\mu_B$, as shown
in Fig. 3(c). The obtained volume fractions for the $\uparrow
\uparrow \downarrow \downarrow$ and $\uparrow \uparrow \uparrow
\downarrow$ phases at 11 T are 10 and 90 \%, respectively. We note
that at 11 T, most of the grains have their $c$ axis oriented
along the field, and therefore the $c$-axis field component is
large enough to stabilize the $\uparrow \uparrow \uparrow
\downarrow$ state in the majority of the grains. This state is
therefore obtained even in a single-phase refinement, albeit with
a small (unphysical) canting angle. Thus, in agreement with the
magnetization data, neutron diffraction also shows that the
$\uparrow \uparrow \uparrow \downarrow$ state is realized at 11 T.
\begin{figure}[htb]
\begin{center}
\epsfig{file=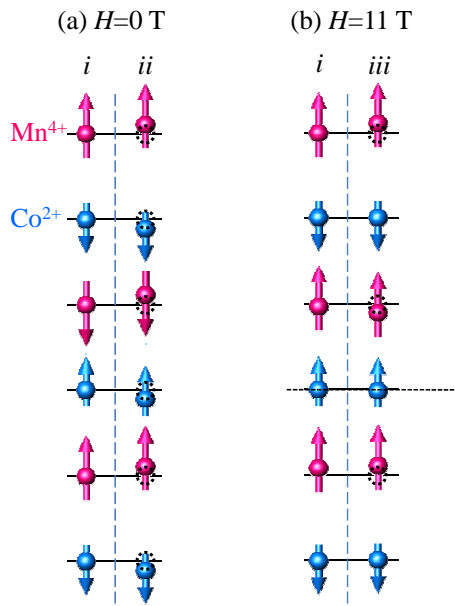, width=6.0 cm} \caption{(color
online) (a) A sketch representing the Ising chains with the
zero-field $\uparrow \uparrow \downarrow \downarrow$ spin order
and alternating ionic charge order, in which electric polarization
is induced through symmetric exchange striction. The corresponding
displacements of the Mn and Co ions along the spin chain are
depicted in (\emph{ii}) relative to their positions (\emph{i}) in
a uniform chain. (b) The $\uparrow \uparrow \uparrow \downarrow$
plateau state without ($i$) and with ($ii$) the ionic
displacements due to the exchange striction. This magnetic and
charge order possesses an inversion symmetry center (horizontal
line).}
\end{center}
\end{figure}
The complete suppression of the electric polarization on the
transition to the $\uparrow \uparrow \uparrow \downarrow$ state
can be easily explained on symmetry grounds. Fig. 4 shows the
$\uparrow \uparrow \downarrow \downarrow$ (a) and $\uparrow
\uparrow \uparrow \downarrow$ (b) undistorted chains ($i$), and
the same chains but with ionic displacements expected as a result
of exchange striction ($ii$ and $iii$). Due to the ionic order in
the chains, collinear electric dipoles form in ($ii$) \cite{choi}.
In contrast, as shown in Fig. 4(b), the $\uparrow \uparrow
\uparrow \downarrow$ state possesses inversion symmetry, and
therefore ($iii$) should lack the polarization, in agreement with
our measurements.

It is well known that interactions between the chains on a
frustrated triangular lattice play a major role in the parent
compound of our series, Ca$_3$Co$_2$O$_6$ \cite{aasland}. Unlike
in our compound, the chains in Ca$_3$Co$_2$O$_6$ are ferromagnetic
and lack intrachain degrees of freedom. Hysteretic magnetization
plateaus are also observed in this compound, but they are
associated with the interchain magnetic order
\cite{hardy2,hardy3}. In contrast, the major features of
Ca$_3$Co$_{2-x}$Mn$_x$O$_6$ as discussed above are dominated by
the physics of a single Ising chain. However, some of our data are
likely to require explanations going beyond this simple model. A
significant hysteresis is observed until the the $\uparrow
\uparrow \uparrow \downarrow$ plateau is achieved (Fig. 2), and a
freezing transition occurs at low temperatures (Fig. 1, and Ref.
\cite{choi}). Frustrated interchain interactions are expected to
play a role for all these effects. The zero-field state is known
to lack true long-range order and to contain finite-size
magnetoelectric domains, probably due to both the frustrated
interchain interactions and the one-dimensionality of this system
\cite{choi}. Gradual magnetization changes, as well as hysteresis,
are expected when a magnetic field is applied and the domains are
gradually converted into the $\uparrow \uparrow \uparrow
\downarrow$ state. The gradual decrease of $P_c$ in an applied
field is probably also associated with these effects. A sharper
magnetization feature may be expected at the transition from the
well-ordered $\uparrow \uparrow \uparrow \downarrow$ state to the
$\uparrow \uparrow \uparrow \uparrow$ state in this Ising system,
as is indeed seen in the data of Fig. 2(a). These effects are
interesting and clearly deserve further investigation. Thus, while
the major features of the phase diagram of
Ca$_3$Co$_{1.05}$Mn$_{0.95}$O$_6$ are established in our work,
further work on this material should be of interest in the context
of the physics of frustrated chain magnets, geometrical
frustration, and magnetically driven ferroelectricity.

In conclusion, Ca$_3$Co$_{2-x}$Mn$_x$O$_6$ ($x\simeq$0.96)
provides a unique experimental realization of a system of
internally frustrated Ising chains assembled on a geometrically
frustrated triangular lattice. Magnetic field induces a transition
from the zero-field $\uparrow \uparrow \downarrow \downarrow$
state characterized by glassiness, to the $\uparrow \uparrow
\uparrow \downarrow$ spin-solid state, and then to the fully
ferromagnetic state. The $\uparrow \uparrow \uparrow \downarrow$
configuration is associated with a magnetization plateau that does
not display any magnetic hysteresis, indicating that this compound
displays a unique interplay between the intra-chain magnetism and
the geometric frustration inherent to its three-dimensional
triangular lattice. This aspect that has yet to be considered
theoretically. We find that a remarkably simple magnetoelectric
model system is realized in this compound. While the $\uparrow
\uparrow \downarrow \downarrow$ state is ferroelectric due to
exchange striction, the field-induced $\uparrow \uparrow \uparrow
\downarrow$ state should possess inversion symmetry. Consistently,
we find that electric polarization is destroyed in the latter
state, giving rise to the observed giant magnetoelectric effect.

The NHMFL is supported by NSF through NSF-DMR-0084173 and the
State of Florida. LB acknowledges the NHMFL in-house research
program and YJJ the NHMFL-Schuller program. Work at Rutgers was
supported by DOE under Grant No. DE-FG02-07ER46382. SL was
partially supported by the Korea Science and Engineering
Foundation through the Center for Strongly Correlated Materials
Research at Seoul National University.


\begin{thebibliography}{}

\bibitem{cheong} S.\ -W.\ Cheong and M.\ Mostovoy, Nature Materials
\textbf{6}, 13 (2007).\

\bibitem{fiebig} M.\ Fiebig, J.\ Phys.\ D: Appl.\ Phys.\ \textbf{38}, R123
(2005).\

\bibitem{khomski} D.\ I.\ Khomskii, J.\ Magn.\ Magn.\ Mater.\ \textbf{306}, 1
(2006).\

\bibitem{mathur} W.\ Eerenstein, N.\ D.\ Mathur, J.\ F.\ Scott, Nature
\textbf{442}, 759 (2006).\

\bibitem{kimura} T.\ Kimura, T.\ Goto, H.\ Shintani, K.\ Ishizaka, T.\ Arima, Y.\ Tokura, Nature \textbf{426}, 55 (2003).\

\bibitem{lawes} G.\ Lawes, A.\ B.\ Harris, T.\ Kimura, N.\ Rogado, R.\ J.\ Cava, A.\ Aharony, O.\ Entin-Wohlman, T.\ Yildirim,
M.\ Kenzelmann, C.\ Broholm, and A.\ P.\ Ramirez, Phys. Rev. Lett. \textbf{95}, 087205
(2005).\

\bibitem{kimura2} T.\ Kimura, J.\ C.\ Lashley, and A.\ P.\ Ramirez, Phys. Rev. B \textbf{73},
220401(R) (2006).\

\bibitem{lee} S.\ Lee, A.\ Pirogov, M.\ S.\ Kang, K.\ H.\ Jang, M.\ Yonemura, T.\ Kamiyama,
S.\ W.\ Cheong, F.\ Gozzo , N.\ Shin, H.\ Kimura, Y.\ Noda , J.\ G.\ Park, Nature \textbf{451}, 805 (2008).\

\bibitem{choi} Y.\ J.\ Choi, H.\ T.\ Yi, S.\ Lee, Q.\ Huang, V.\ Kiryukhin, and S.\-W.\ Cheong, Phys.\ Rev.\ Lett.\ \textbf{100},
047601 (2008).\

\bibitem{subkov} V.\ G.\ Zubkov, G.\ V.\ Bazuev, A.\ P.\ Tyutyunnik, I.\ F.\ Berger, J.\ Solid State Chem.\ \textbf{160}, 293
(2001).\

\bibitem{selke} W.\ Selke, Phys. Rep. \textbf{170}, 213 (1998).\

\bibitem{jjkim} J.\ -J.\ Kim, S. Mori, I. Harada, J.\ Phys.\ Soc.\ Japan
\textbf{65}, 2624 (1996).\

\bibitem{fullprof} J.\ Rodriguez-Carvajal, Physica B \textbf{192},
55 (1993).\

\bibitem{ueda} H.\ Ueda  H.\ A.\ Katori, H.\ Mitamura, T.\ Goto, and H.\ Takagi, Phys.\ Rev.\ Lett.\ \textbf{94},
047202 (2005); H.\ Ueda, H.\ Mitamura, T.\ Goto, and Y.\ Ueda,
Phys.\ Rev.\ B \textbf{73}, 094415 (2006).\

\bibitem{aasland} S.\ Aasland, H.\ Fjellvag, B.\ Hauback, Solid State Commun.
\textbf{101}, 187 (1997).\

\bibitem{hardy2} V.\ Hardy, M.\ R.\ Lees, O.\ A.\ Petrenko, D.\ McK.\ Paul, D.\ Flahaut, S.\ Hébert, and A.\ Maignan, Phys.\ Rev.\ B
\textbf{70}, 064424 (2004).\

\bibitem{hardy3} V.\ Hardy, D.\ Flahaut, M.\ R.\ Lees, O.\ A.\ Petrenko,  Phys.\ Rev.\ B \textbf{70},
214439 (2004).\


\end{thebibliography}

\end{document}